\documentclass[twocolumn,aps,amsmath,amssymb,floatfix,superscriptaddress,prb,longbibliography]{revtex4-1}
\usepackage{graphicx}
\usepackage{dcolumn}
\usepackage{bm}
\UseRawInputEncoding
\usepackage{xspace}
\usepackage{hyperref}
\usepackage{color}
\usepackage[flushleft]{threeparttable}
\def\new{\color{black}}
\def\NCTO{Na$_2$Cu$_2$TeO$_6$\xspace}
\begin{document}

\title{Evidence for strong correlations at finite temperatures in the dimerized magnet Na$_2$Cu$_2$TeO$_6$}

\author{Yanyan~Shangguan}
\altaffiliation{These authors contributed equally to the work.}
\author{Song~Bao}
\altaffiliation{These authors contributed equally to the work.}
\affiliation{National Laboratory of Solid State Microstructures and Department of Physics, Nanjing University, Nanjing 210093, China}
\author{Zhao-Yang~Dong}
\altaffiliation{These authors contributed equally to the work.}
\affiliation{Department of Applied Physics, Nanjing University of Science and Technology, Nanjing 210094, China}
\author{Zhengwei Cai}
\affiliation{National Laboratory of Solid State Microstructures and Department of Physics, Nanjing University, Nanjing 210093, China}
\author{Wei~Wang}
\affiliation{School of Science, Nanjing University of Posts and Telecommunications (NUPT), Nanjing 210023, China}
\author{Zhentao~Huang}
\affiliation{National Laboratory of Solid State Microstructures and Department of Physics, Nanjing University, Nanjing 210093, China}
\author{Zhen~Ma}
\affiliation{Institute for Advanced Materials, Hubei Normal University, Huangshi 435002, China}
\author{Junbo~Liao}
\affiliation{National Laboratory of Solid State Microstructures and Department of Physics, Nanjing University, Nanjing 210093, China}
\author{Xiaoxue~Zhao}
\affiliation{National Laboratory of Solid State Microstructures and Department of Physics, Nanjing University, Nanjing 210093, China}
\author{Ryoichi Kajimoto}
\affiliation{J-PARC Center, Japan Atomic Energy Agency (JAEA), Tokai, Ibaraki 319-1195, Japan}
\author{Kazuki~Iida}
\affiliation{Neutron Science and Technology Center, Comprehensive Research Organization for Science and Society (CROSS), Tokai, Ibaraki 319-1106, Japan}
\author{David~Voneshen}
\affiliation{ISIS Facility, Rutherford Appleton Laboratory, Chilton, Didcot, Oxon, OX11 0QX, United Kingdom}
\affiliation{Department of Physics, Royal Holloway University of London, Egham, TW20 0EX, United Kingdom}
\author{Shun-Li~Yu}
\email{slyu@nju.edu.cn}
\author{Jian-Xin~Li}
\email{jxli@nju.edu.cn}
\author{Jinsheng~Wen}
\email{jwen@nju.edu.cn}
\affiliation{National Laboratory of Solid State Microstructures and Department of Physics, Nanjing University, Nanjing 210093, China}
\affiliation{Collaborative Innovation Center of Advanced Microstructures, Nanjing University, Nanjing 210093, China}

\begin{abstract}

Dimerized magnets forming alternating Heisenberg chains exhibit quantum coherence and entanglement and thus can find potential applications in quantum information and computation. However, magnetic systems typically undergo thermal decoherence at finite temperatures. Here, we show inelastic neutron scattering results on an alternating antiferromagnetic-ferromagnetic chain compound Na$_2$Cu$_2$TeO$_6$
that the excited quasiparticles can counter thermal decoherence and maintain strong correlations at elevated temperatures. At low temperatures, we observe clear dispersive singlet-triplet excitations arising from the dimers formed along the crystalline $b$-axis. The excitation gap is of $\sim$18~meV and the bandwidth is about half of the gap. The band top energy has a weak modulation along the [100] direction, indicative of a small interchain coupling. The gap increases while the bandwidth decreases with increasing temperature, leading to a strong reduction in the available phase space for the triplons. As a result, the Lorentzian-type energy broadening becomes highly asymmetric as the temperature is raised. These results are associated with a strongly correlated state resulting from hard-core constraint and quasiparticle interactions. We consider these results to be not only evidence for strong correlations at finite temperatures in Na$_2$Cu$_2$TeO$_6$, but also for the universality of the strongly correlated state in a broad range of quantum magnetic systems.

\end{abstract}

\maketitle

\section{Introduction}

Spin-chain systems confined to one dimension provide an excellent simple platform in studying the quantum magnetism\cite{0953-8984-1-19-001}. In particular, spin-chain compounds with Heisenberg interactions alternating between the intradimers and interdimers have been studied extensively by means of inelastic neutron scattering (INS)\cite{PhysRevB.30.6300,doi:10.1063/1.340736,PhysRevB.38.543,
PhysRevB.46.8268,PhysRevB.50.9174,PhysRevB.53.15004,PhysRevB.54.R9624,
PhysRevB.54.R6827,PhysRevB.63.172414,PhysRevLett.90.227204,PhysRevB.67.054414,
PhysRevB.69.104417,PhysRevB.98.104413}. These materials have an $S=0$ singlet ground state, in which two neighboring antiparallel spins within a dimer bind together to form a quantum mechanical singlet, and effectively the system does not exhibit long-range order even down to zero temperature. The spectra feature a singlet-triplet excitation separated by a gap $\Delta$, the energy required to flip a spin within a singlet. The triplet excitations correspond to the triply degenerate spin $S=1$  quasiparticles termed triplons which may exhibit topological properties due to the presence of Dzyaloshinskii-Moriya interactions or competing exchange couplings\cite{PhysRevLett.84.5876,doi:10.1143/JPSJ.74.2189,
PhysRevLett.93.267202,nc6_6805,np12_224,np13_736,nc10_2096}.
The Hamiltonian for an alternating Heisenberg chain system can be written as\cite{PhysRevB.59.11384,PhysRevLett.84.4465,PhysRevB.98.104413},
\begin{equation}\label{eq:hamiltonian}
H=\sum_{i=1,N/2}J_1{\bm S}_{2i-1}\cdot{\bm S}_{2i}+J_2{\bm S}_{2i}\cdot{\bm S}_{2i+1},
\end{equation}
where $N$ is the total number of spins in a chain, $J_1$ and $J_2$ are the interactions for the spins within a dimer and between the nearest-neighbor dimers, respectively, and $\eta=|J_2/J_1|$ which is typically less than 1 is the alternating parameter. A finite $J_2$ will give rise to the dispersive excitation spectra. To the lowest order of the interdimer interactions, the dispersion can be simplified as\cite{PhysRevB.59.11384,PhysRevLett.84.4465,PhysRevB.98.104413}
\begin{equation}\label{eq:dispersion}
E({\bm Q})=J_1-\frac{J_2}{2}\cos({\bm Q}d),
\end{equation}
where ${\bm Q}$ is the wave vector and $d$ is the distance between the nearest-neighbor dimers (from center to center).  In this case, $\Delta$ and the bandwidth $B$ can be approximated to be $J_1-|J_2|/2$ and $|J_2|$, respectively. This indicates the critical role of $\eta$ in controlling the available phase space for the triplons.

\begin{figure}[htb]
\centerline{\includegraphics[width=3.in]{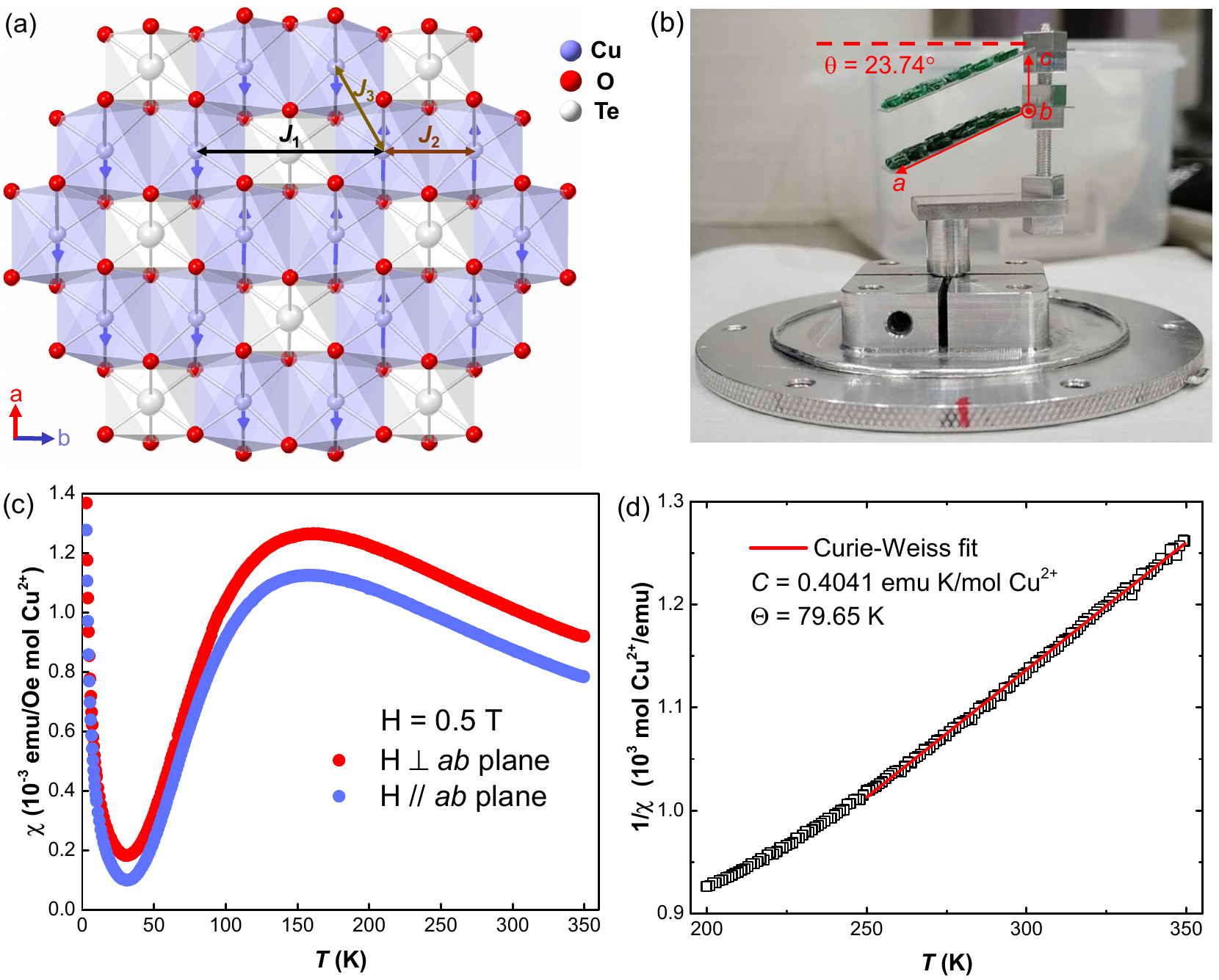}}
\caption{(a) Schematic Cu$_2$TeO$_6$ layer structure of \NCTO. Arrows denote the spins. $J_1$, $J_2$, and $J_3$ represent the intradimer (antiferromagnetic), interdimer (ferromagnetic), and interchain (ferromagnetic) couplings, respectively. (b) Coaligned single crystals we used for the INS measurements. This photo only shows 60 pieces of single crystals, whereas in the actual measurements we used 90 pieces glued on 3 aluminum plates. The aluminum plates were well tilted by 23.74$^\circ$ so that the horizontal plane was the ($H,\,K,\,0$) plane. (c) Temperature dependence of the magnetic susceptibility measured with magnetic field of $\mu_0H$ = 0.5~T applied parallel and perpendicular to the $a$-$b$ plane. {\new There is some anisotropy of the spin susceptibility, which may be due to the quasi-one-dimensional magnetic structure and impurity spins.} (d) Inverse magnetic susceptibility of \NCTO between 200 and 350~K. The solid line represents a fit to the data with the Curie-Weiss law over the interval from 250 to 350~K, yielding the parameters indicated in the figure.
\label{fig1}}
\end{figure}

{\new Recently, a dimerized magnet \NCTO was proposed to be a quasi-one-dimensional chain system\cite{xu2005synthesis,doi:10.1143/JPSJ.75.083709,PhysRevB.76.104403,inorgchem47_128,PhysRevB.89.174403,C4CE01382D,PhysRevB.102.220402}.
It has a monoclinic structure with the space group $C2/m$, and consists of alternating Cu$_2$TeO$_6$ magnetic layers and Na nonmagnetic layers stacking along the $c$-axis in a monoclinic way\cite{xu2005synthesis}, similar to Na$_3$Cu$_2$SbO$_6$~(Refs.~\onlinecite{miura2006spin,doi:10.1143/JPSJ.77.104709}). In the Cu$_2$TeO$_6$ layer, six CuO$_6$ octahedra form a distorted honeycomb lattice via edge-sharing with a TeO$_6$ octahedron at the center as shown in Fig.~\ref{fig1}(a). Each Cu$^{2+}$ with spin-1/2 interacts with three neighboring spins [Fig.~\ref{fig1}(a)], but it has been suggested that the third-neighbor interchain coupling ($J_3$) can be ignored and only the antiferromagnetic (AFM) $J_1$ and ferromagnetic (FM) $J_2$ along the $b$-axis dominate, making \NCTO a quasi-one-dimensional alternating AFM-FM system\cite{xu2005synthesis,doi:10.1143/JPSJ.75.083709,PhysRevB.76.104403,inorgchem47_128,PhysRevB.89.174403}.
Interestingly, the magnitude of $J_1$ is larger than that of $J_2$ although it involves a super-superexchange path Cu-O-Te-O-Cu which is much longer\cite{xu2005synthesis}. Very recently, Gao {\it et al.} have performed INS measurements on a single crystal of \NCTO, which reveal clear singlet-triplet excitations\cite{PhysRevB.102.220402}. In Ref.~\onlinecite{PhysRevB.102.220402}, they show the excitation spectra with a gap of $\sim$18~meV can be described by a Hamiltonian with $J_1=22.78(2)$~meV, $J_2=-8.73(4)$~meV, and a small but non-negligible antiferromagnetic $J_3=1.34(3)$~meV. These results place \NCTO in a unique position\cite{PhysRevB.102.220402}. First, the large gap and small bandwidth in \NCTO make the available phase space for the triplons severely limited. Second, the interactions alternate antiferromagnetically and ferromagnetically along the chain.  Third, it has a small but finite interchain coupling, which preserves the weak two-dimensional nature of the magnetic interactions. One intriguing property of the dimerized magnets with limited phase space is that a strongly correlated state at high temperatures may be retained due to the hard-core constraint (only one quasiparticle is allowed per dimer even for bosons) and quasiparticle interactions\cite{PhysRevB.78.094411,PhysRevB.78.100403,Essler_2009,PhysRevB.82.104417,PhysRevB.85.014402,PhysRevLett.109.127206,PhysRevB.89.134407,PhysRevB.90.024428,PhysRevB.93.241109}. The strong correlations at finite temperatures, which can be characterized by the asymmetric Lorentzian-type broadening of the energy scans instead of the symmetric Lorentzian-type broadening as in conventional magnets, have been observed in spin chains\cite{PhysRevB.85.014402,PhysRevB.90.024428,PhysRevB.93.134404,PhysRevB.93.241109,PhysRevB.98.104435}, spin ladders\cite{PhysRevLett.96.047210,PhysRevLett.100.157204}, and three-dimensional gapped quantum magnet Sr$_3$Cr$_2$O$_8$~(Refs.~\onlinecite{PhysRevLett.109.127206,PhysRevB.89.134407}), but not in two-dimensional AFM-FM alternating chain compounds yet. Considering these\cite{xu2005synthesis,PhysRevB.102.220402}, \NCTO provides an ideal platform  to investigate the universality of the strongly correlated state at elevated temperatures in quantum magnets.}

{\new Here, we report comprehensive results of INS measurements on \NCTO and focus on the temperature dependence of the spectra. Our results reveal that \NCTO is an alternating AFM-FM chain compound with weak but non-negligible interchain coupling, in excellent agreement with a previous study\cite{PhysRevB.102.220402}. Interestingly, by following the temperature dependence of the excitation spectra, we find that the gap increases while the bandwidth decreases with increasing temperature, thus limiting the available phase space for the triplons. In the meantime, we observe an asymmetric Lorentzian-type energy broadening, which gets more pronounced as the temperature is raised.} We attribute this to the thermally activated strongly correlated state arising from hard-core constraint and quasiparticles interactions. Our results on the asymmetric Lorentzian-type energy broadenings are compelling evidence for the strong correlations at finite temperatures in \NCTO, and indicate the universal presence of strongly correlated state in a broad range of quantum systems.

\section{Experimental Details}

Polycrystalline powders of \NCTO were synthesized by a standard solid-state reaction method, and single crystals were successfully grown by the flux method as described in Ref.~\onlinecite{C4CE01382D}. The grown translucent green crystals with $a$-$b$ plane as the cleavage plane have a typical mass of 30~mg for each piece. Magnetic susceptibility measurements were conducted using the vibrating sample magnetometer option integrated in a Physical Property Measurement System (PPMS-9T) from Quantum Design.

For INS experiments, single crystals were glued onto aluminum plates by hydrogen-free Cytop grease. These crystals were well coaligned using a backscattering Laue x-ray diffractometer. Neutron scattering experiments were performed on time-of-flight spectrometers MERLIN at ISIS facility in the United Kingdom\cite{bewley2009merlin} and 4SEASONS at J-PARC center in Japan\cite{doi:10.1143/JPSJS.80SB.SB025}. We coaligned 90 pieces of single crystals weighed about 3~g in total for the experiments on MERLIN and 4SEASONS. As shown in Fig.~\ref{fig1}(b), single crystals were coaligned onto the rectangular aluminum plates with $a$ and $b$ axes oriented along the edge directions of the aluminum plate. These crystals were well aligned so that the overall mosaic spread was $\sim$0.3$^{\circ}$ as determined from the rocking scan through the (2,\,0,\,0) peak. The aluminum plates were well tilted by 23.74$^{\circ}$ so that the horizontal plane was the ($H,\,K,\,0$) plane. The assembly with ($H,\,K,\,0$) as the horizontal plane was mounted in a closed-cycle refrigerator for neutron experiments.

For the measurements on MERLIN, we set the angle of the neutron beam direction parallel to the $b$-axis to be zero. Data were collected at 5 and 18~K with $E_{\rm i}=40.49$~meV by rotating the sample about the vertical direction with a range of 120$^\circ$ in a 2$^\circ$ step. As for the measurements on 4SEASONS, we used a primary $E_{\rm i}= 55$~meV and a chopper frequency of 350~Hz with an energy resolution of 2.65~meV at the elastic line. Measurements were performed at 5, 90, 150 and 200~K. We set the angle where the [100] direction was parallel to the incident beam direction to be 0$^{\circ}$. Data were collected by rotating the sample about the vertical axis from -30$^{\circ}$ to 150$^{\circ}$ in a 1$^{\circ}$ step. We counted 15 minutes for each step. Raw data were reduced and analyzed using Horace\cite{EWINGS2016132}. The wave vector $\boldsymbol Q$ was expressed as ($H,\,K,\,L$) reciprocal lattice unit~(rlu) of $(a^*,\,b^*,\,c^*)$ = ($2\pi/a\cos\theta,\,2\pi/b,\, 2\pi/c\cos\theta$) with $a = 5.706$~\AA, $b = 8.6375$~\AA, $c = 5.938$~\AA, and $\theta=113.74^{\circ}-90^{\circ}=23.74^{\circ}$, representing the angle between $a(c)$ and $a^*(c^*)$.

\section{Results}

{\new\subsection{Sample characterizations and magnetic excitation spectra}}

We measured the magnetic susceptibility of the \NCTO single crystals with a vibrating sample magnetometer, and the results are shown in Fig.~\ref{fig1}(c) and (d). From Fig.~\ref{fig1}(c), the magnetic susceptibility does not show any signature of long-range order. There is a broad hump with its maximum around 150~K, which indicates
the gradual establishment of short-range singlets below this temperature, and thus the susceptibility decreases. The hump signifies the existence of a spin gap. By fitting the susceptibility with an alternating AFM-FM chain model\cite{ic20_1033,ic33_5171,PhysRevB.59.11384,PhysRevB.76.104403,
inorgchem47_128,PhysRevB.89.174403}, a gap of $\Delta\sim$ 254~K was estimated for \NCTO~(Refs.~\onlinecite{xu2005synthesis,miura2006spin}). {\new The upturn at low temperatures is probably due to the impurity spins, which together with the quasi-one-dimensional magnetic structure may cause the anisotropy of the susceptibility with field applied in and out of the plane.} Examination of the high-temperature data of the inverse magnetic susceptibility [Fig.~\ref{fig1}(d)] shows that $\chi$ follows the Curie-Weiss law, with $\chi= C/(T+\Theta)$, from 250 to 350~K. From the Curie-Weiss fit, we obtain the Curie constant, $C = 0.4041$~emu~K/mol~Cu$^{2+}$, which is consistent with a spin-only effective magnetic moment $\mu_{\rm{eff}}=1.80$~$\mu_{\rm{B}}$/Cu$^{2+}$ with $S = 1/2$. This leads to a Land\'e $g$-factor of 2.08 for this compound, which is within the reasonable range for Cu$^{2+}$. The Curie-Weiss temperature $\Theta$ is 79.65~K, with the positive sign indicative of the dominant antiferromagnetic correlation. These results are in excellent agreement with previous studies on this compound\cite{xu2005synthesis,miura2006spin,PhysRevB.76.104403,
inorgchem47_128,PhysRevB.89.174403}.

\begin{figure*}[htb]
\centerline{\includegraphics[width=5.6in]{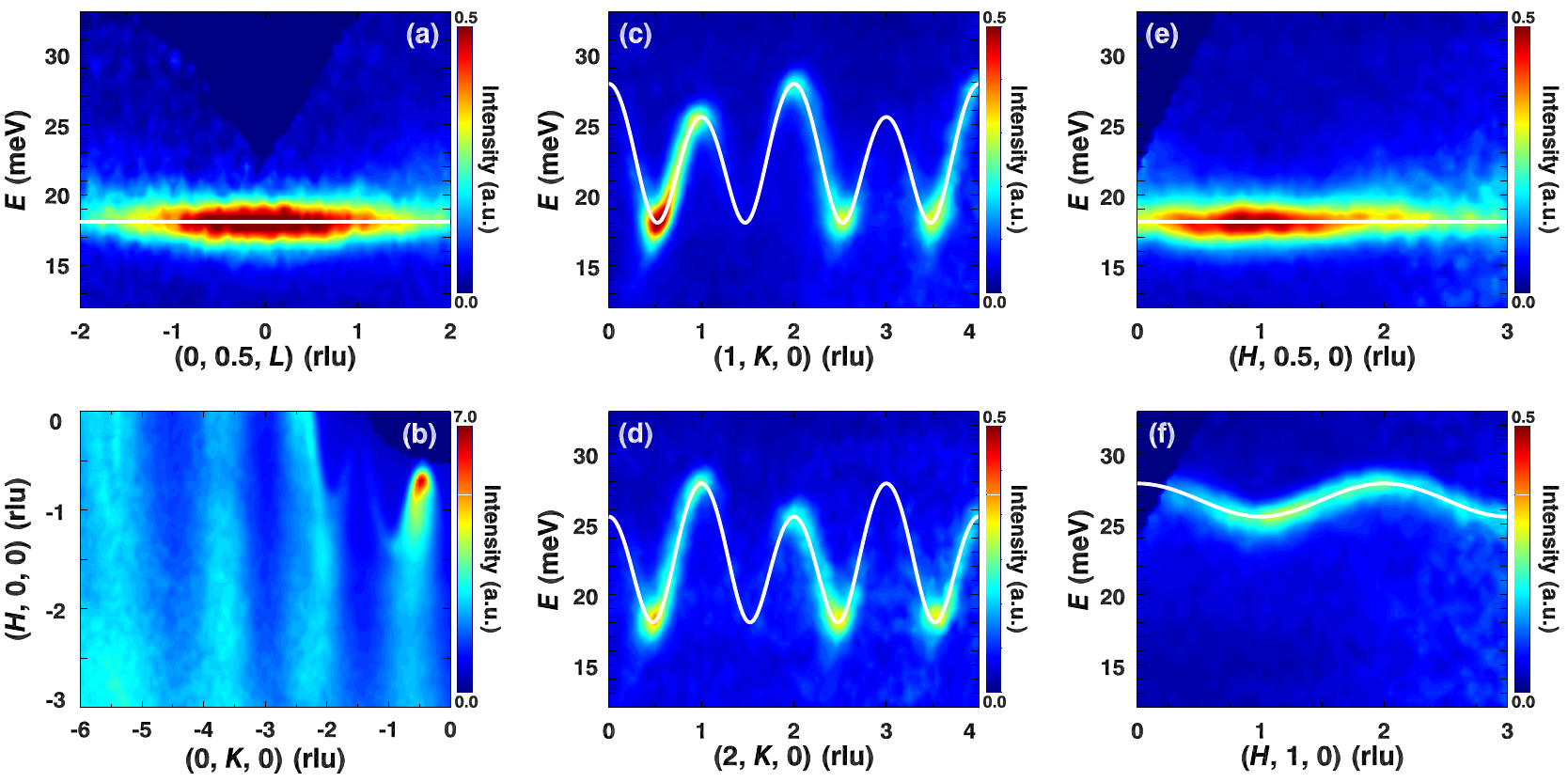}}
\caption{Magnetic spectra of \NCTO measured at 5~K. (a) Scattering intensity measured along ($0,\,0.5,\,L$) with integration widths of $\Delta H$ and $\Delta K$ of 0.2~rlu. (b) Contour map of the constant-energy cut measured with $E_{\rm i}= 40.49$~meV, obtained on MERLIN at ISIS. The energy range of the integration is [15,30]~meV. (c)-(f) INS spectra collected along ($1,\,K,\,0$), ($2,\,K,\,0$), ($H,\,0.5,\,0$) and ($H,\,1,\,0$) within the ($H,\,K,\,0$) plane with $E_{\rm i}= 55$~meV. The integration widths are $\Delta H$ = 0.2~rlu, $\Delta K$ = 0.2~rlu, and $\Delta L$ = 4~rlu. {\new Solid curves are the calculated dispersions using exchange parameters of $J_1=22.82(7)$~meV, $J_2=-8.47(8)$~meV, and $J_3=1.37(6)$~meV.} The data except (b) are from 4SEASONS at J-PARC.
\label{fig2}}
\end{figure*}

We have performed INS measurements on the \NCTO single crystals, and the singlet-triplet  excitation spectra along the $H$, $K$, and $L$ directions at $T= 5$~K are plotted in Fig.~\ref{fig2}. {\new These magnetic excitation spectra share a close similarity with those presented in Ref.~\onlinecite{PhysRevB.102.220402}. To quantify the magnetic correlations of \NCTO and compare the extracted exchange coupling parameters with those in the previous study\cite{PhysRevB.102.220402}, random phase approximation calculations of the magnetic excitation spectra were performed. We first extracted the experimental dispersions from modified Lorentzian fits to constant-{$\bm Q$} scans at 145 points along five high-symmetry directions, which were then fitted by the same dispersion relation utilized in Ref.~\onlinecite{PhysRevB.102.220402}. The fitted parameters resulting from our fits are $J_1=22.82(7)$~meV, $J_2=-8.47(8)$~meV, and $J_3=1.37(6)$~meV, which are almost the same as those of Ref.~\onlinecite{PhysRevB.102.220402}. The calculated dispersions using these parameters are plotted on top of the experimental data as solid curves in Fig.~\ref{fig2}. The calculated dispersions are in excellent agreement with the experimental results. Our theoretical calculations confirm the weakly two-dimensional nature of the magnetic interactions in \NCTO, where the dimers are coupled to form alternating chains along the $b$-axis and the interchain coupling within the $a$-$b$ plane is weak but non-negligible\cite{PhysRevB.102.220402}. Based on these magnetic excitation spectra, we explore their temperature evolution in details in the following.}

\begin{figure*}[htb]
\centerline{\includegraphics[width=6.in]{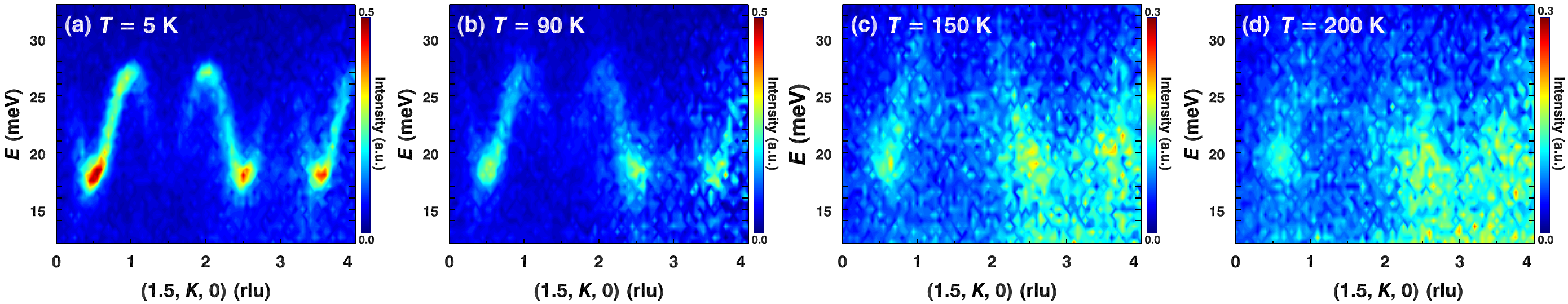}}
\caption{Temperature dependence of the magnetic excitation spectra. (a)-(d) INS spectra along ($1.5,\,K,\,0$) at $T$ = 5, 90, 150 and 200~K, respectively, collected on 4SEASONS at J-PARC. The integration widths are $\Delta H$ = 0.2~rlu, and $\Delta L$ = 4~rlu.
\label{fig3}}
\end{figure*}

\subsection{Strong correlated behaviors}

\begin{figure}[htb]
\centerline{\includegraphics[width=3.in]{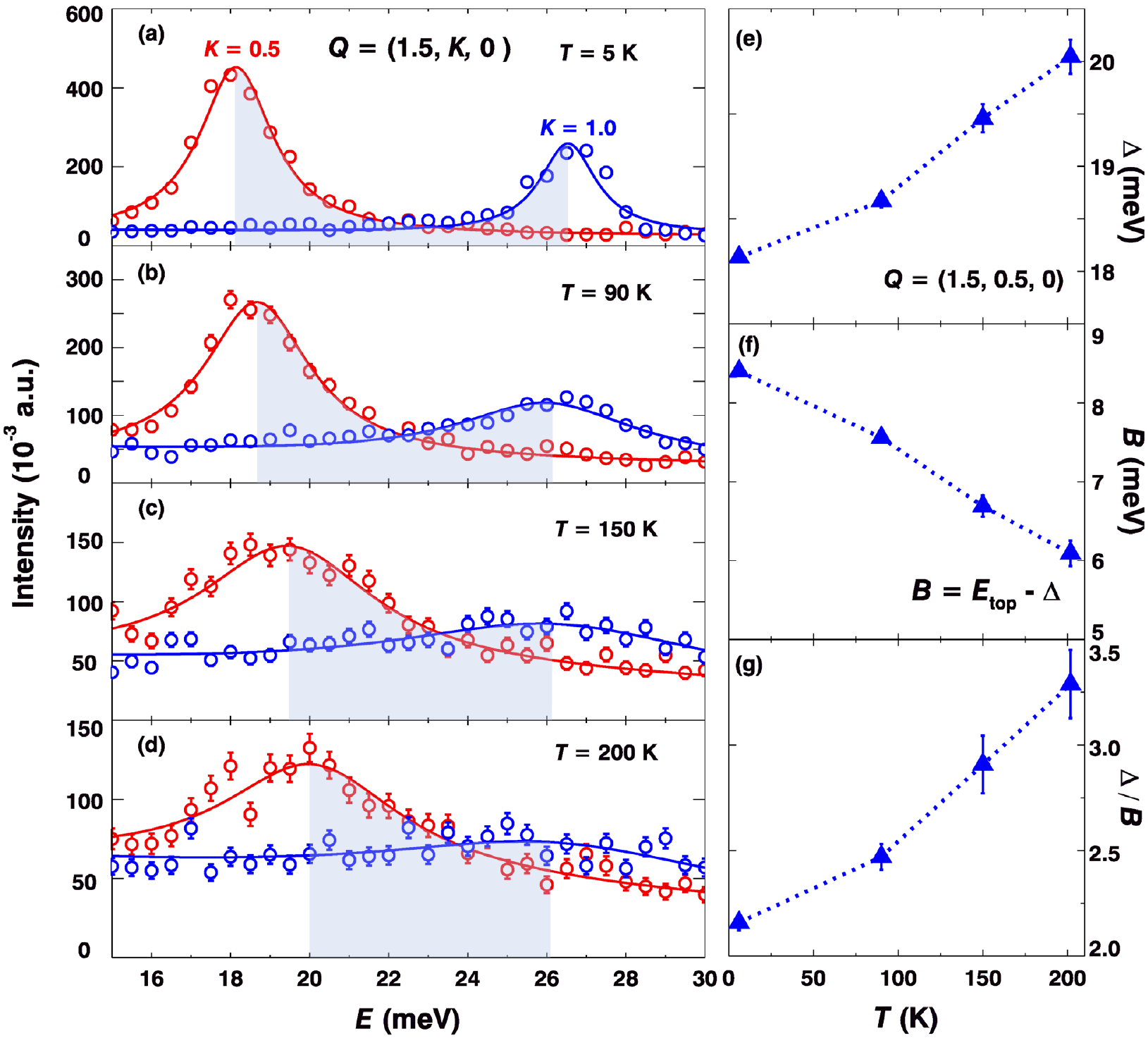}}
\caption{(a)-(d) Constant-{$\bm Q$} cuts upon the spectra of Fig.~\ref{fig3} at the band bottom (1.5,\,0.5,\,0) and band top (1.5,\,1.0,\,0) at 5, 90, 150 and 200~K, respectively. The integration intervals for $H$ and $K$ are both $\pm$0.1~rlu, while for $L$ it is $\pm$2~rlu. Solid lines correspond to the fits using Lorentzian function from which we extract the peak positions. The shades denote the bandwidth $B$ which is the energy interval between the centers of the band bottom and top. (e)-(g) Temperature dependence of gap energy $\Delta$, bandwidth $B$ and the ratio of spin gap to bandwidth $\Delta$/$B$, respectively.
\label{fig4}}
\end{figure}

\begin{figure}[htb]
\centerline{\includegraphics[width=2.8in]{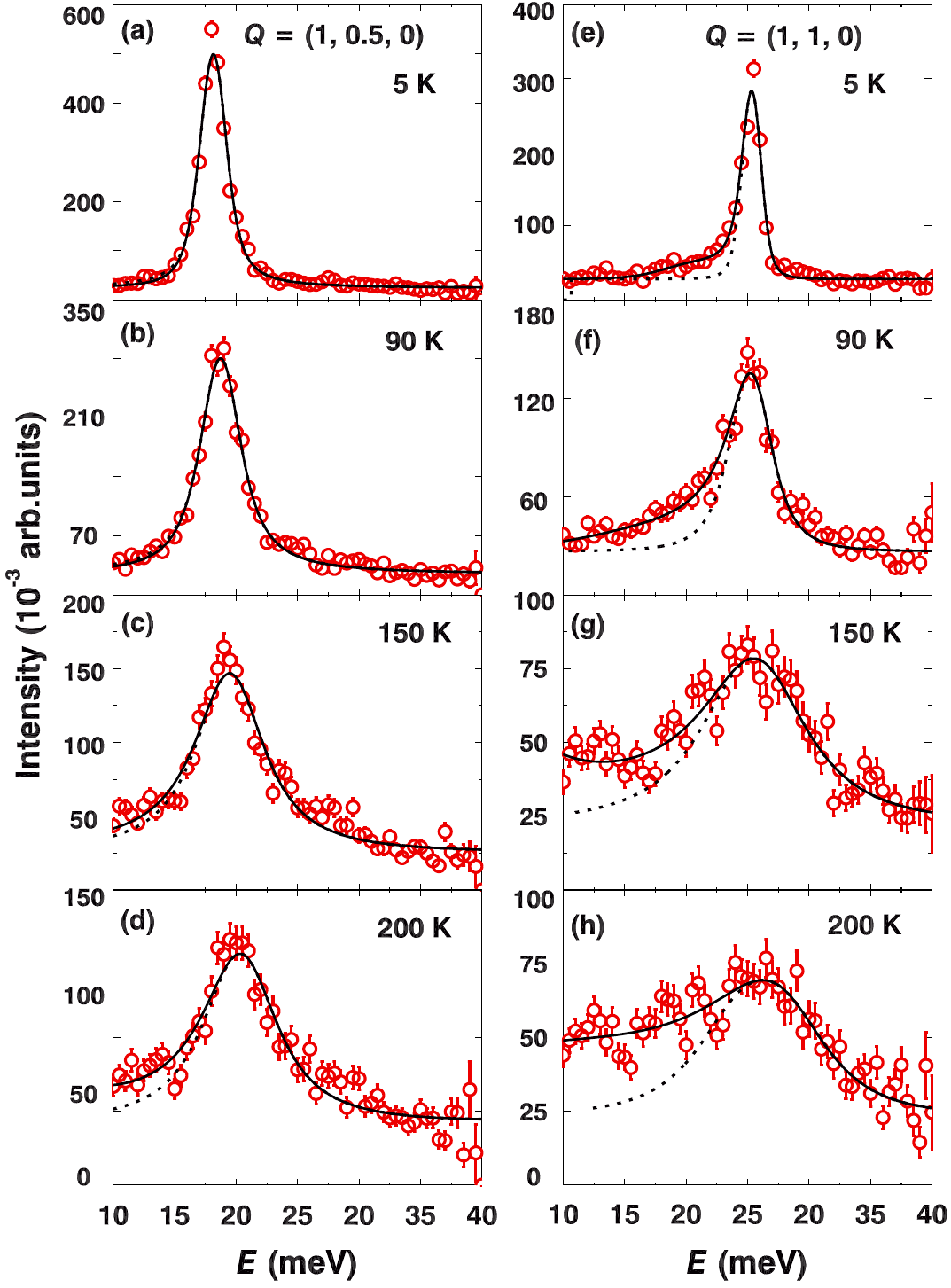}}
\caption{(a)-(d) Constant-$\bm Q$ cuts at the band bottom (1,\,0.5,\,0), and (e)-(h) at the band top (1,\,1,\,0) at 5, 90, 150 and 200~K, obtained on 4SEASONS at J-PARC. The integration intervals for $H$ and $K$ are both $\pm$0.1~rlu, while for $L$ it is $\pm$2~rlu. Black solid lines correspond to the fits using Eq.~(\ref{eq:al}). {\new Dashed curves represent the symmetric Lorentzian fits.}
\label{fig5}}
\end{figure}

\begin{figure}[htb]
\centerline{\includegraphics[width=2.8in]{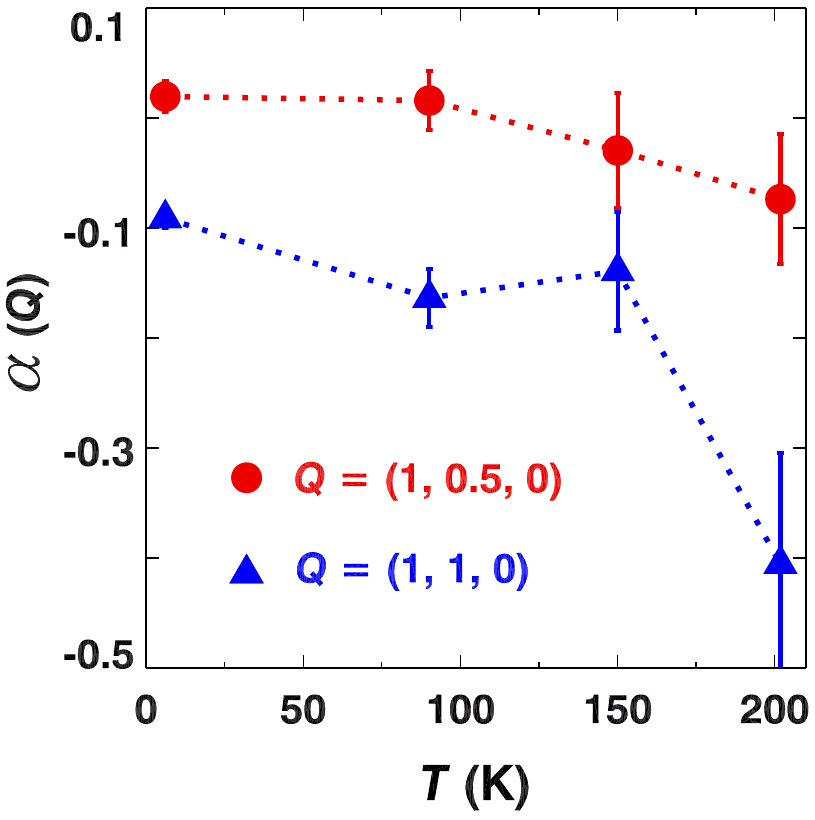}}
\caption{Temperature dependence of the asymmetry factor $\alpha({\bm Q})$ extracted from fits with Eq.~(\ref{eq:al}) to the data in Fig.~\ref{fig5}.
\label{fig6}}
\end{figure}

The magnetic excitation spectra along (1.5,\,$K$,\,0) at 5, 90, 150, and 200~K are plotted in Fig.~\ref{fig3}. We can clearly observe the well-defined triplon excitations at low temperatures [Fig.~\ref{fig3}(a)]. At 90~K, which is below the hump temperature, the triplon bands are still clearly resolvable, although the intensity becomes much weaker compared to that at 5~K. At 150 and 200~K, the excitations are much broader, but the excitations around $K=0.5$, 2.5, and 3.5~rlu can still be identified. To further quantify the triplon band, we plot constant-${\bm Q}$ scans at the band bottom (1.5,\,0.5,\,0) and band top (1.5,\,1,\,0) at different temperatures in Fig.~\ref{fig4}(a)-(d). As the temperature is raised, the scattering intensity gets weaker, and the peak width, especially for that at (1.5,\,1,\,0), becomes much larger. The temperature-induced intensity weakening and peak broadening are expected for a magnetic system. From Fig.~\ref{fig4}(a)-(d), we can extract the spin gap $\Delta$, bandwidth $B$, and the ratio of spin gap to bandwidth $\Delta$/$B$ at different temperatures. We take the peak center of the band bottom to be $\Delta$, and the energy interval between the centers of the band bottom and top to be $B$ as we illustrate by the shades in Fig.~\ref{fig4}(a)-(d).

These results are plotted in Fig.~\ref{fig4}(e)-(g). In Fig.~\ref{fig4}(e), the gap increases monotonically with increasing temperature at ${\bm Q}=(1.5,\,0.5,\,0)$ obtained on 4SEASONS. This behavior is different from that of a magnetically ordered system where the gap size decreases with increasing temperature due to the reduced ordered magnetic moment. On the other hand, it is similar to the Haldane-chain compounds with integer spins, where the gap increases with temperature as a result of the reduced coherent length and finite-size effect\cite{PhysRevB.33.659,PhysRevB.50.9265}. The bandwidth, as presented in Fig.~\ref{fig4}(f), decreases monotonically with temperature. This can also be visualized from Fig.~\ref{fig4}(a)-(d) by the shades, where the band bottom increases while the band top decreases from 5 to 200~K. As a consequence, the ratio of the gap to the bandwidth increases from 2.16 at 5~K to 3.29 at 200~K, as shown in Fig.~\ref{fig4}(g). This ratio quantifies the available phase space for the triplons. A larger ratio results in a more limited available phase space, and hence the triplons will be localized and interacting strongly with each other. This is verified by analyzing the line shape of the constant-$\boldsymbol Q$ cuts at (1,\,0.5,\,0) and (1,\,1,\,0) at different temperatures in Fig.~\ref{fig5}.

From Fig.~\ref{fig5}, it is clear that the line shapes of the excitations do not simply broaden symmetrically with increasing temperature. At (1,\,0.5,\,0) at 5~K, the energy scan at the band bottom is sharp and symmetric~[Fig.~\ref{fig5}(a)]. As the temperature is increased, the peak becomes asymmetric with the tail extending towards lower energies. At (1,\,1,\,0), where the mode is at the top of the excitation band, the temperature-induced asymmetry is more clearly revealed~[Fig.~\ref{fig5}(e)-(h)]. In fact, there are two broadening mechanisms of the excitation in our experiment. One is the Gaussian profile originating from the instrument resolution and mosaic spread of the sample, which is independent of temperature. The other one is the symmetric Lorentzian profile resulting from the finite lifetime of magnons, {\new as illustrated by the dashed lines in Fig.~\ref{fig5}. However, these two mechanisms will not give rise to the highly-asymmetric line shapes. As the scans in Fig.~\ref{fig5} are at relatively small ${\bm Q}$s with $H=1$~rlu, and $K=0.5$ and 1~rlu, phonon contributions are not significant (for $H=1$~rlu, phonon contributions only become noticeable for $K>3$~rlu). Furthermore, if phonons were to contribute to the scattering, the line shapes would become broad symmetrically. As for the impurity spins, they should have strong effect at low temperatures only as shown in Fig.~\ref{fig1}(c), and will not give rise to the asymmetric line shapes which are only significant at high temperatures either. Therefore, the asymmetric line shapes must be an intrinsic property of \NCTO related to the Cu spins. As in previous studies\cite{PhysRevB.85.014402,PhysRevLett.109.127206,PhysRevB.93.134404,PhysRevB.93.241109}, to account for the asymmetry, we use a function consisting of a modified Lorentzian convoluted with a Gaussian function, which represents the instrument resolution, to fit our data:}
\begin{equation}\label{eq:al}
\begin{aligned}
&&I({\bm Q},E)&=A({\bm Q})\int_{-\infty}^{\infty}dt\frac{\exp\{-\frac{[E-\Gamma({\bm Q})t-E_0({\bm Q})]^2}{2\sigma^2(E)}\}}{\sqrt{2\pi\sigma^2(E)}}\\
&&      &\times\{\frac{1}{\pi}\frac{1}{1+[t-\alpha({\bm Q})t^2+\beta({\bm Q})t^3]^2}+c{_0}\}.
\end{aligned}
\end{equation}
Here, $A({\bm Q})$ is the overall amplitude, $t$ represents $[E^\prime-E_0(\bm Q)]/\Gamma(\bm Q)$ [$\Gamma(\bm Q)$ is the half width at half maximum of the Lorentzian function], $E_0({\bm Q})$ is the center of the peak, and $\sigma$ is the variance of the Guassian indicative of the instrument resolution. $\beta({\bm Q})$ is the damping term and $c_0$ is a constant. The asymmetry is controlled by $\alpha({\bm Q})$, and the sign of $\alpha({\bm Q})$ governs whether the peak broadens to higher [$\alpha({\bm Q})>0$] or lower energies [$\alpha({\bm Q})<0$]. When $\alpha({\bm Q}),\,\beta({\bm Q})=0$, the modified Lorentzian profile becomes a regular symmetric Lorentzian. The data can be fitted reasonably well with Eq.~(\ref{eq:al}) as shown in Fig.~\ref{fig5}. In fact, we have also tried to fit the data with the modified Lorentzian function only without convoluting the resolution (not shown) and compared the fitting curves and output parameters by the two approaches. We find that there is no much difference between them except that the instrument resolution has a marginal effect on the broadening of the line shape. We extract the temperature dependence of the asymmetry indicator $\alpha({\bm Q})$, and plot it in Fig.~\ref{fig6}. The value of $\alpha({\bm Q})$ is much larger at (1,\,1,\,0) than that at (1,\,0.5,\,0), indicating a much larger asymmetry of the line shape as can be visualized from Fig.~\ref{fig5}. At both ${\bm Q}$s, $\alpha({\bm Q})$ increases with temperature, which is in excellent agreement with the more pronounced asymmetric Lorentzian-type line shape in Fig.~\ref{fig5}. At high temperatures, the asymmetry parameter at (1,\,0.5,\,0) and (1,\,1,\,0) are both negative, indicating that the peak broadens to lower energies. 

\section{Discussions}

{\new Our INS results on the spin dynamics of \NCTO characterize it as an alternating AFM-FM chain compound with a weak but finite interchain coupling and a limited phase space due to the large $\Delta/B$. Notably, by mapping the magnetic spectra at elevated temperatures combined with the constant-$\boldsymbol Q$ cuts, we find that the available phase space for the triplons is more limited at high temperatures. Meanwhile, we observe an asymmetric Lorentzian-type energy broadening, which gets more pronounced as the temperature is raised shown in Fig.~\ref{fig5}. Furthermore, we quantitatively parameterize the temperature evolution of the asymmetric energy broadening by fitting the constant-$\boldsymbol Q$ cuts to an asymmetric Lorentzian function convoluted with the instrument resolution. The asymmetric parameter $\alpha({\bm Q})$ increases with increasing temperature. Our results on the asymmetric Lorentzian-type energy broadenings reflect the intrinsic properties of the Cu spins and are compelling evidence for the strong correlations at finite temperatures in \NCTO~(Refs.~\onlinecite{PhysRevB.85.014402,PhysRevB.90.024428,PhysRevB.93.134404,PhysRevB.93.241109,PhysRevB.98.104435,PhysRevLett.96.047210,PhysRevLett.100.157204,PhysRevLett.109.127206,PhysRevB.89.134407}). Considering the weak two-dimensionality of the magnetic interactions and the alternating antiferromagnetic-ferromagnetic couplings along the chain in \NCTO, our results indicate the universal presence of strongly correlated states in a broad range of quantum systems, beyond spin chains\cite{PhysRevB.85.014402,PhysRevB.90.024428,PhysRevB.93.134404,PhysRevB.93.241109,PhysRevB.98.104435}, spin ladders\cite{PhysRevLett.96.047210,PhysRevLett.100.157204}, as well as three-dimensional quantum magnets\cite{PhysRevLett.109.127206,PhysRevB.89.134407}.}




We believe the microscopic origin of the asymmetric line shape and the strong correlation state at finite temperatures is the hard-core constraint and small available phase space for quasiparticle scattering\cite{PhysRevB.78.094411,PhysRevB.78.100403,Essler_2009,PhysRevB.82.104417,PhysRevB.85.014402,PhysRevLett.109.127206,PhysRevB.89.134407,PhysRevB.90.024428,PhysRevB.93.241109}. For dimerized magnets, the excitations are spin-1 triplons created by exciting a dimer from a singlet to a triplet. These quasiparticles are subject to hard-core constraint, with which they cannot occupy the same dimer site. At finite temperatures, a significant number of dimer triplets are thermally activated. In the case that the available space is limited with quasi-one-dimensionality and large $\Delta/B$, the neutron-excited triplons will interact strongly with the nearby thermally-activated triplons due to the hard-core constraint. The combined effect of a smaller phase space and larger thermal population of triplons with increasing temperature will give rise to a more asymmetric line shape as described in the text for \NCTO due to the hard-core constraint, as is also the case for other dimerized systems\cite{PhysRevB.85.014402,PhysRevB.90.024428,PhysRevB.93.134404,PhysRevB.93.241109,PhysRevB.98.104435,PhysRevLett.96.047210,PhysRevLett.100.157204,PhysRevLett.109.127206,PhysRevB.89.134407}. {\new We notice that the asymmetry has a strong ${\bm Q}$-dependence, which should be related to the interaction strength and the density of states of the triplons\cite{PhysRevLett.109.127206,PhysRevB.93.241109}.} Because of the strong correlation effect at high temperatures, random phase approximation calculations may not describe the high-temperature data well as they ignore correlations between the triplons\cite{PhysRevB.85.014402,PhysRevLett.109.127206,PhysRevB.93.134404,PhysRevLett.84.4465,PhysRevB.102.220402}.

{\new\section{CONCLUSIONS}}

{\new Our comprehensive INS excitation spectra clearly identify \NCTO to be an alternating AFM-FM chain compound with weak but non-zero interchain coupling. More importantly, we find the energy scans show asymmetric line shapes at elevated temperatures different from conventional magnets. The asymmetry becomes more pronounced accompanying the reduced phase space for the triplons and the increasing population of temperature-activated triplons as the temperature is raised. We take it as compelling evidence that in a weakly two-dimensional alternating AFM-FM dimer magnet \NCTO, the triplons which are subject to the hard-core constraint interact more strongly at high temperatures. These results indicate the presence of strongly correlated state in a broad range of quantum magnetic systems. Such systems with the capability of countering thermal decoherence may find future applications in quantum information and computation\cite{RevModPhys.80.517,nielsen2002quantum}.}

\section{ACKNOWLEDGMENTS}
We would like to thank Rui~Chen and Prof.~Junfeng~Wang at Wuhan National High Magnetic Field Center for assisting us in measuring the magnetization under high magnetic fields. The work was supported by National Key Projects for Research and Development of China with Grant No.~2021YFA1400400, National Natural Science Foundation of China with Grant Nos.~11822405, 12074174, 12074175, 11774152, 11904170, and 12004191, Natural Science Foundation of Jiangsu province with Grant Nos.~BK20180006, BK20190436, and BK20200738, Hubei Provincial Natural Science Foundation of China with Grant No.~2021CFB238, Fundamental Research Funds for the Central Universities with Grant No.~020414380183, and the Office of International Cooperation and Exchanges of Nanjing University. We acknowledge the neutron beam time from ISIS with Proposal No.~2010131 and from J-PARC with Proposal No.~2020A0037. Z.M. thanks Beijing National Laboratory for Condensed Matter Physics for funding support.


%

\end{document}